\documentclass[useAMS,usenatbib]{mn2e}
\usepackage{color}
\usepackage{graphicx}
\usepackage{graphics}
\usepackage{rotating}
\usepackage{amsmath}        
\usepackage{amssymb}        
\usepackage{subfigure}        
\usepackage{epsfig}        
\usepackage{longtable}
\usepackage[T1]{fontenc}
\usepackage{aecompl}

\newcommand{\mnras}{MNRAS}
\newcommand{\apjl}{ApJ}
\newcommand{\apj}{ApJ}
\newcommand{\apjs}{ApJS}
\newcommand{\aj}{AJ}
\newcommand{\nat}{Nature}
\newcommand{\aap}{A\&A}

\newcommand{\pasj}{PASJ}

\newcommand{\pasp}{PASP}

\newcommand{\nar}{NAR}
\newcommand{\physrep}{PR}

\newcommand{\source}{SS~Cyg}

\newcommand{\perbeam}{\,beam$^{-1}$}

\newcommand{\Swift}{{\it Swift}}

\title[Radio outbursts of \source{}]{The reproducible radio outbursts of SS Cygni}

\author[T.~D.~Russell et al.]{T.~D.~Russell,$^{1}$\thanks{email: t.russell@curtin.edu.au} J.~C.~A.~Miller-Jones,$^{1}$ G.~R.~Sivakoff,$^{2}$ D.~Altamirano,$^{3}$ \and T.~J.~O'Brien,$^{4}$ K.~L.~Page,$^{5}$ M.~R.~Templeton, E.~G.~K\"ording,$^{6}$ C.~Knigge,$^3$ \and M.~P.~Rupen,$^{7}$ R.~P.~Fender,$^{8}$ S.~Heinz,$^{9}$ D.~Maitra,$^{10}$ S.~Markoff,$^{11}$ S.~Migliari,$^{12,13}$ \and R.~A.~Remillard,$^{14}$ D.~M.~Russell,$^{15}$ C.~L.~Sarazin,$^{16}$ and E.~O.~Waagen$^{17}$\\
$^1$International Centre for Radio Astronomy Research - Curtin University, GPO Box U1987, Perth, WA 6845, Australia\\
$^{2}$Department of Physics, University of Alberta, 4-181 CCIS, Edmonton, AB T6G 2E1, Canada\\
$^3$Department of Physics and Astronomy, University of Southampton, Highfield SO17 IBJ, England\\
$^{4}$Jodrell Bank Centre for Astrophysics, University of Manchester, Manchester M13 9PL, UK\\
$^{5}$Department of Physics and Astronomy, University of Leicester, Leicester, LE1 7RH, UK\\
$^{6}$Department of Astrophysics/IMAPP, Radboud University Nijmegen, PO Box 9010, NL-6500 GL Nijmegen, the Netherlands \\
$^{7}$National Research Council, Herzberg Astronomy and Astrophysics, 717 White Lake Road, PO Box 248, Penticton, \\British Columbia V2A 6J9, Canada\\
$^{8}$Department of Physics, Oxford University, Denys Wilkinson Building, Keble Road, Oxford OX1 3RH, UK \\
$^{9}$Astronomy Department, University of Wisconsin-Madison, 475. N. Charter St., Madison, WI 53706, USA \\
$^{10}$Department of Physics and Astronomy, Wheaton College, Norton, MA 02766, USA \\
$^{11}$Anton Pannekoek Institute for Astronomy, University of Amsterdam, P.O. Box 94249, 1090 GE Amsterdam, the Netherlands \\
$^{12}$European Space Astronomy Centre (ESAC/ESA), Camino Bajo del Castillo s/n, Urb. Villafranca del Castillo, 28692, Villanueva \\ de la Ca\~{n}ada, Madrid, Spain. \\
$^{13}$Department of Astronomy and Meteorology \& Institute of Cosmic Sciences, University of Barcelona, Mart\'{i} i Franqu\`{e}s 1, \\ 08028 Barcelona, Spain. \\
$^{14}$MIT Kavli Institute for Astrophysics and Space Research, Building 37, 70 Vassar Street, Cambridge, MA 02139, USA \\
$^{15}$New York University Abu Dhabi, P.O. Box 129188, Abu Dhabi, United Arab Emirates \\
$^{16}$Department of Astronomy, University of Virginia, P.O. Box 400325, Charlottesville, VA 22904, USA \\
$^{17}$American Association of Variable Star Observers, 49 Bay State Road, Cambridge, MA 02138\\
}

\begin{document}
\date{Accepted 2016 May 19.}
\pagerange{\pageref{firstpage}--\pageref{lastpage}} \pubyear{2015}
\maketitle

\label{firstpage}
\begin{abstract}
We present the results of our intensive radio observing campaign of the dwarf nova \source{} during its 2010 April outburst. We argue that the observed radio emission was produced by synchrotron emission from a transient radio jet. Comparing the radio light curves from previous and subsequent outbursts of this system (including high-resolution observations from outbursts in 2011 and 2012) shows that the typical long and short outbursts of this system exhibit reproducible radio outbursts that do not vary significantly between outbursts, which is consistent with the similarity of the observed optical, ultraviolet and X-ray light curves. Contemporaneous optical and X-ray observations show that the radio emission appears to have been triggered at the same time as the initial X-ray flare, which occurs as disk material first reaches the boundary layer. This raises the possibility that the boundary region may be involved in jet production in accreting white dwarf systems. Our high spatial resolution monitoring shows that the compact jet remained active throughout the outburst with no radio quenching.

\end{abstract}

\begin{keywords}
radio continuum: stars -- X-rays: stars -- stars: jets, cataclysmic variables, individual (\source{})
\end{keywords}

\section{Introduction}

Cataclysmic variables (CVs) are a class of interacting binary system comprised of a white dwarf that is accreting matter from a red dwarf companion via Roche lobe overflow \citep[see][for a comprehensive review of these systems]{1995CAS....28.....W}.  Dwarf novae are a subclass of CVs in which the K or M dwarf donor star fills its Roche lobe and the transferred matter forms an accretion disk around the weakly-magnetised white dwarf (magnetic field $B\lesssim10^6$\,G; e.g. \citealt{1995CAS....28.....W}, referred to as non-magnetic). Accretion on to the white dwarf occurs via a boundary layer between the rapidly spinning inner accretion disk and the more slowly spinning white dwarf. These systems undergo episodic outbursts, with typical recurrence timescales varying between $\sim10$\,days and several decades. The outburst mechanism is generally explained by the disk instability model (e.g. \citealt{1974PASJ...26..429O, 1979PThPh..61.1307H, 1982ApJ...260L..83C, 1983MNRAS.205..359F}, see \citealt[][]{2001NewAR..45..449L} for a thorough treatment). In this model, the disk viscosity is believed to be too low during quiescence to transport the accreted matter steadily through the disk and onto the white dwarf surface. Therefore, the disk density gradually increases up to a critical point whereupon it becomes optically thick, the temperature and viscosity increase, and mass is rapidly transferred through the disk onto the white dwarf, causing an outburst. The outburst duration is believed to depend on the mass present in the disk at the onset of the instability \citep{1993ApJ...419..318C}, and the rise and decay times correspond to the duration over which the heating or cooling front propagates \citep{2001NewAR..45..449L}. A source may exhibit alternating long and short outbursts \citep[e.g.][]{1992ApJ...401..642C} due to the effects of tidal dissipation and heating of the outer disk by the impact of the accretion stream, as well as the mass transfer rate fluctuating in a small range about a critical value \citep{2001A&A...366..612B}.

With a degenerate compact object accreting matter from a less-evolved donor star, dwarf novae show many similarities to low-mass X-ray binaries (LMXBs). However, while the existence of jets has been confirmed in certain accretion states of both black hole and neutron star X-ray binaries with the exception of the high-magnetic field neutron star systems (see \citealt{2006csxs.book..381F} and \citealt{2010LNP...794...85G} for a review), few CVs have been well studied at radio wavelengths. Until recently, it was believed that dwarf novae, and CVs in general, did not launch jets, placing strong constraints on the jet-launching mechanism \citep[e.g.][]{1999PhR...311..225L, 2004A&A...422.1039S}. 

Given the prevalence of jets in all other known classes of object with accretion disks \citep[see, e.g.,][]{1999PhR...311..225L}, the absence of jets in dwarf novae was surprising. Good evidence for the existence of jets from accreting white dwarfs has been found in at least ten of the $\sim$200 known symbiotic systems \citep{2004MNRAS.347..430B, 2004RMxAC..20...35S}, which are systems consisting of a white dwarf accreting material from a red giant companion star. However, these systems are not believed to contain accretion disks. Accretion is instead thought to proceed via Bondi-Hoyle capture, with the observed outbursts usually being attributed to a thermonuclear runaway of the accreted material on the surface of the white dwarf. During the 2006 nova explosion of the symbiotic recurrent nova RS~Oph, highly-collimated bipolar jets ending in moving synchrotron-emitting lobes were directly resolved by radio imaging \citep{2006Natur.442..279O, 2008ApJ...685L.137S}.  Although the outbursts of recurrent novae such as RS~Oph are traditionally explained as nuclear explosions on the white dwarf surface, \citet{2009MNRAS.397L..51K} instead suggest that this system is in fact a dwarf nova with a very large accretion disk, which could then provide a mechanism for launching and collimating the observed jets. Transient jets have also been observed in super soft X-ray sources \citep[e.g.][]{1996ApJ...456..320C, 1998ApJ...504..854C, 1998A&A...338L..13M, 1998ApJ...506..880B}. These very luminous objects contain a highly-accreting white dwarf thought to be undergoing steady nuclear burning \citep[e.g.][]{1992A&A...262...97V}. However, the steady nuclear burning requires mass transfer rates at least 10 times higher than than those occurring in CVs, and occurs via thermally-unstable Roche lobe overflow, where the donor star is more massive than the accreting object.    

Radio observations of the dwarf nova \source{} during its 2007 April outburst revealed variable radio emission, consistent with a partially self-absorbed jet \citep{2008Sci...320.1318K}. Radio observations of subsequent \source{} outbursts have also shown radio flaring during outburst, followed by fainter, steady emission, with no radio quenching during the outburst \citep{2011IAUS..275..224M, 2013Sci...340..950M}. Recently, four non-magnetic nova-like systems, which are CVs with mass transfer rates sufficiently high that their accretion disk is maintained in a constant hot state, have also been detected at radio wavelengths \citep{2011MNRAS.418L.129K, 2015MNRAS.451.3801C}. However, the radio emission mechanism for nova-like systems remains unclear, with possibilities including synchrotron, gyrosynchrotron or coherent emission \citep{2015MNRAS.451.3801C}. 

During a typical outburst, X-ray binaries evolve through a range of characteristic accretion states \citep[e.g.][]{2011BASI...39..409B}, defined by specific X-ray spectral and variability behaviour. The power and morphology of the jets in these systems is very well coupled with the observed X-ray state in both black hole \citep[e.g.][]{2004MNRAS.355.1105F} and neutron star \citep[e.g.][]{2006MNRAS.366...79M} systems, implying a fundamental coupling between inflow in the accretion disk and outflow in the jets. Specifically, in black hole systems the hard X-ray states seen at the beginning and end of an outburst are associated with steady, slightly inverted or flat-spectrum ($\alpha$$\gtrsim$0, where the source flux density $S_{\nu}$ varies with frequency $\nu$ as, $S_{\nu} \propto \nu^{\alpha}$), compact jets \citep[e.g.][]{2000A&A...359..251C, 2000ApJ...543..373D, 2001MNRAS.327.1273S}, whereas bright, optically-thin ($\alpha < 0$), relativistically-moving ejecta are seen at transitions to the soft states \citep[e.g.][]{1994Natur.371...46M, 2004ApJ...617.1272C, 2004MNRAS.355.1105F, 2012MNRAS.421..468M}. Low-magnetic field neutron star X-ray binaries trace out similar X-ray evolution patterns to their black hole counterparts \citep{2004ApJ...608..444M}, and the disk-jet coupling is also believed to be broadly similar in these systems \citep{2006MNRAS.366...79M}, with radio emission being triggered at transitions between hard and soft states just as in the black hole systems \citep{2009MNRAS.400.2111T, 2010ApJ...716L.109M}.

\subsection{\source{}}

\source{} is the prototypical dwarf nova, located at a distance of 114$\pm$2\,pc \citep{2013Sci...340..950M, 2013ApJ...773L..26N}. This object has a very well-sampled optical light curve \citep{1992ApJ...401..642C}  stretching back to its discovery in 1896 \citep{1896HarCi..12....1P}.  The long-term behaviour of \source{} shows an outburst recurrence time of $49\pm15$\,d and a bimodal distribution of outburst durations, with peaks at 7 and 14\,d \citep{1992ApJ...401..642C}. During outburst, the visual magnitude rises from the quiescent visual magnitude of $m_{\rm vis}\sim$12 to $m_{\rm vis}\sim$8.5. Despite the bimodal distribution of outburst durations, this system displays a narrow range of outburst rise and decay times, exhibiting a rise of $0.56\pm0.14$\,d\,mag$^{-1}$ and a decay rate of $2.38\pm0.27$\,d\,mag$^{-1}$ \citep{1998ApJ...505..344C}. A small fraction ($\sim$17\%; \citealt{1998ApJ...505..344C}) of outbursts are observed to be slow-rise (historically referred to as ``anomalous''), which rise at a rate of between 1.25 and 3.75\,d\,mag$^{-1}$, and do not appear to be strongly peaked around one particular rise time \citep{1998ApJ...505..344C}.

The most detailed ultraviolet and X-ray coverage (probing the inner regions of the accretion flow) of an outburst from \source{} was provided by \citet{2003MNRAS.345...49W}, using data from the {\it Extreme Ultraviolet Explorer} ({\it EUVE}) and the {\it Rossi X-ray Timing Explorer} ({\it RXTE}). These observations showed that the initial optical rise was accompanied by a hard X-ray flare, which rapidly quenched as the boundary layer became optically thick to its own radiation and cooled efficiently, causing the hard X-rays to give way to the extreme ultraviolet emission. Residual hard X-ray emission persisted through the outburst phase (fainter than quiescent levels) before rising again, more slowly this time, at the end of the optical and ultraviolet decay phase, before eventually fading back to quiescent levels.

Following up on the detection of radio emission during outburst \citep{2008Sci...320.1318K}, \source{} was targeted with a multi-wavelength monitoring campaign during its 2010 April outburst. Triggered from monitoring data provided by the American Association of Variable Star Observers (AAVSO), the 2010 campaign included radio observations taken with the Very Large Array (VLA), the Westerbork Synthesis Radio Telescope (WSRT) the Very Long Baseline Array (VLBA) and the European Very long baseline interferometry Network (EVN) at radio wavelengths, AAVSO in the optical band, as well as {\it RXTE} and {\it Swift} at X-ray wavelengths.  In Section \ref{sec:observations} we describe this observing campaign and our data analysis methods.  We detail our results in Section \ref{sec:results}, and go on to discuss the nature of the radio emission and implications for the jet-disk coupling in Section \ref{sec:jd_coupling}, before summarising our conclusions in Section \ref{sec:conclusions}.

\section{Observations}
\label{sec:observations}

\subsection{VLA}

Following notification from the AAVSO in 2010 April that a new outburst of \source{} was underway, we triggered radio observations during the commissioning phase of the Karl G. Jansky Very Large Array (VLA) and were able to get on source within 24\,h of the alert (project code AM991).  Our initial trigger observation was made with the X-band system, using 256\,MHz of contiguous bandwidth centered at 8.46\,GHz.  Following the initial radio detection, we switched the observing band to C-band, observing in two 128-MHz sub-bands, each made up of 64 channels of width 2\,MHz, centred at 4.6 and 7.9\,GHz.  We obtained 9 epochs of VLA data over the course of 19 days before the outburst finished (determined by AAVSO monitoring).  The observations are summarised in Table \ref{tab:evla}.  Throughout the observing campaign, the array was in its most compact D-configuration.

\begin{table}
\caption{VLA observations of \source{}. Flux density errors are statistical uncertainties on the fitted source parameters (systematic uncertainties of 1\% have not been included). Specified Modified Julian Dates (MJDs) are for the mid-point of the observation, with the quoted uncertainty of $\pm$0.01\,days reflecting the observation duration.}
\centering
\label{tab:evla}
\begin{tabular}{cccc}
\hline
Date & MJD & Frequency & Flux density\\
& $\pm$0.01 & (GHz) & (mJy)\\
\hline
2010 Apr 21 & 55307.46 & 8.459$\pm$0.128 & 0.20$\pm$0.03 \\
2010 Apr 22 & 55308.47 & 4.599$\pm$0.064 & 0.81$\pm$0.03 \\
	    & 	       & 7.899$\pm$0.064 & 0.72$\pm$0.03 \\
2010 Apr 23 & 55309.47 & 4.599$\pm$0.064 & 0.62$\pm$0.03 \\ 
	    &	       & 7.899$\pm$0.064 & 0.41$\pm$0.02 \\
2010 Apr 25 & 55311.59 & 4.599$\pm$0.064 & 0.28$\pm$0.02 \\
	    &	       & 7.899$\pm$0.064 & 0.25$\pm$0.02 \\
2010 Apr 27 & 55313.35 & 4.599$\pm$0.064 & 0.52$\pm$0.04 \\
	    &	       & 7.899$\pm$0.064 & 0.44$\pm$0.04 \\
2010 Apr 30 & 55316.47 & 4.599$\pm$0.064 & 0.21$\pm$0.02 \\
	    & 	       & 7.899$\pm$0.064 & 0.20$\pm$0.02 \\
2010 May 02 & 55318.42 & 4.599$\pm$0.064 & 0.28$\pm$0.02 \\
	    & 	       & 7.899$\pm$0.064 & 0.20$\pm$0.02 \\
2010 May 06 & 55322.49 & 4.599$\pm$0.064 & 0.16$\pm$0.02 \\
	    & 	       & 7.899$\pm$0.064 & 0.18$\pm$0.02 \\
2010 May 09 & 55325.47 & 4.599$\pm$0.064 & 0.18$\pm$0.02 \\
	    &          & 7.899$\pm$0.064 & 0.14$\pm$0.02 \\
\hline
\end{tabular}
\end{table}

\begin{figure}
\centering
\includegraphics[width=\columnwidth]{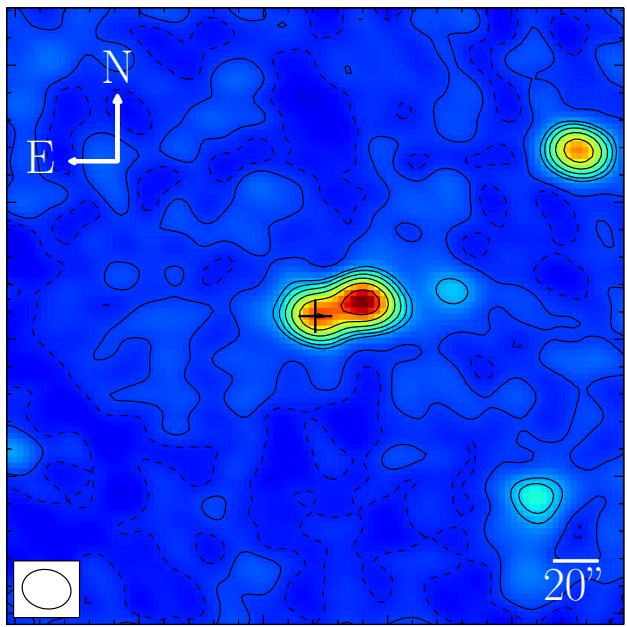}
\caption{2010 April 22 (MJD 55308) 4.6\,GHz VLA image of \source{}, showing the 1.16$\pm$0.05\,mJy (at 4.6\,GHz) confusing source 22\farcs3 to the WNW of the target at a position of 21:42:40.932$\pm$0$\farcs$16 in R.A. and 43:35:16.281$\pm$0$\farcs$12 in Declination. \source{} is marked by the VLBA source position{} (black cross). Contour levels are at $\pm \left( \sqrt{2}^{n}  \right)$ times the rms noise (25\,$\mu$Jy), where $n=-3,3,4,5,6,...$ (dashed contours represent negative values). We estimate the spectral index of the nearby source to be $\alpha=-0.05\pm0.09$. }
\label{fig:nearbySource}
\end{figure}

The data were first written to uvfits format within the Common Astronomy Software Application (CASA; \citealt{2007ASPC..376..127M}) before we carried out the bulk of the data reduction within the Astronomical Image Processing System (AIPS; \citealt{2003ASSL..285..109G}). The secondary calibrator J2202+4216 was used to determine the delays and then calibrate the bandpass. We used J1331+3030 (3C286) to calibrate the overall flux density scale according to the coefficients derived at the VLA in 2010. Amplitude and phase gains were transferred from the secondary calibrator to the target source, and data were averaged in frequency within each sub-band prior to imaging. Flux densities were measured by fitting a point source to the target in the image plane.

There were a number of additional sources within the field, particularly in the 4.6\,GHz sub-band. The presence of a 1.16-mJy source (at 4.6\,GHz) only 22\farcs3 to the WNW of \source{} (Figure~\ref{fig:nearbySource}) required us to subtract this source in the {\it uv}-plane prior to fitting for the flux density of \source{}. The nearby source did not vary between observations, has a fitted position of 21:42:40.932$\pm$0$\farcs$16 in R.A. and 43:35:16.281$\pm$0$\farcs$12 in Declination and a spectral index of $\alpha$=$-$0.05$\pm$0.09.

\subsection{WSRT}

The Westerbork Synthesis Radio Telescope (WSRT) observations were carried out on 2010 April 24 (09:39--13:07 UT), April 26 (05:57--13:00 UT) and April 28 (04:31--05:43 UT). We observed at a median frequency of 4.901\,GHz in full polarization mode with eight 20-MHz sub-bands, each comprising 64 channels, for a total bandwidth of 160\,MHz. Standard flagging and calibration, as well as imaging were done in MIRIAD \citep{1995ASPC...77..433S}. The reported flux density of the point source was measured in the image-plane (Table~\ref{tab:wsrt}).

The observation taken on April 28 was only 70\,min long and resulted in a very poor quality image (WSRT is an east-west array). Therefore, we were unable to derive any useful information from this observation.

\begin{table}
\caption{WSRT observations of \source{}, observed at 4.901\,GHz with a bandwidth of 160\,MHz. Flux density errors are statistical uncertainties on the fitted source parameters. Specified Modified Julian Dates (MJDs) are for the mid-point of the observation, with the quoted uncertainties reflecting the observation duration.}
\centering
\label{tab:wsrt}
\begin{tabular}{ccc}
\hline
Date & MJD &  Flux density\\
& & (mJy)\\
\hline
2010 Apr 24 & 55310.48$\pm$0.06  & 0.36$\pm$0.03 \\
2010 Apr 26 & 55312.39$\pm$0.15  & 0.20$\pm$0.03 \\
\hline
\end{tabular}
\end{table}

\subsection{VLBA}

Having detected \source{} in the radio band with the VLA, we triggered higher angular resolution observations with the VLBA under project code BM308. We observed at 8.4\,GHz in dual circular polarization, using the maximum available recording rate of 512\,Mbps, corresponding to an observing bandwidth of 64\,MHz per polarization. The observations were phase-referenced to the nearby calibrator source J2136+4301, from the fifth VLBA Calibrator Survey \citep[VCS-5;][]{2007AJ....133.1236K} and located 1.3$^{\circ}$ from \source{}.  We switched between target and calibrator with a cycle time of 3\,min, substituting the VCS-3 \citep{2005AJ....129.1163P} check source J2153+4322 for every seventh scan on the target.  At the start and end of every observing run, we observed a range of bright calibrator sources at differing elevations to better solve for un-modelled clock and tropospheric phase errors using the AIPS task DELZN, thereby improving the success of the phase transfer.  Data reduction was carried out according to standard procedures within AIPS. Flux densities were calculated by fitting a point source to the target in the image plane (Table~\ref{tab:vlba}). These observations have previously been reported by \citet{2013Sci...340..950M}. 

We have also included VLBA observations from 2012 August 12 and 2012 October 30 (project code BS215), as reported by \citet{2013Sci...340..950M}.

\begin{table}
\caption{8.4\,GHz  VLBA observations of \source{}. Specified Modified Julian Dates (MJDs) are for the mid-point of the observation, with the quoted uncertainties reflecting the observation duration. Systematic flux uncertainties of 5\% have not been included.}
\centering
\label{tab:vlba}
\begin{tabular}{ccc}
\hline
Date & MJD & Flux density\\
& & (mJy)\\
\hline
2010 Apr 22 & 55308.58$\pm$0.14 & 0.47$\pm$0.05 \\
2010 Apr 23 & 55309.58$\pm$0.14 & 0.39$\pm$0.06 \\
2010 Apr 24 & 55311.76$\pm$0.13 & 0.29$\pm$0.06{$^{\rm a}$} \\
2010 Apr 30 & 55316.77$\pm$0.10 & 0.39$\pm$0.07 \\
2010 May 02 & 55318.43$\pm$0.04 & 0.25$\pm$0.07 \\
2010 May 06 & 55322.74$\pm$0.13 & $\leq$0.15 \\

2012 Aug 12 & 56151.29$\pm$0.12 & 0.93$\pm$0.06{$^{\rm b}$} \\
2012 Oct 30 & 56231.10$\pm$0.11 & 0.17$\pm$0.05{$^{\rm b}$} \\

\hline
\multicolumn{3}{l}{$^{\rm a}$ Flux density measured for the core of \source{} only, this}\\
\multicolumn{3}{l}{\phantom{$^{\rm a}$} does not include the possible jet component (Section~\ref{sec:resolved}).}\\
\multicolumn{3}{l}{$^{\rm b}$ 8.4\,GHz VLBA results from \citet{2013Sci...340..950M}}\\

\end{tabular}
\end{table}

\subsection{EVN}
We include two observations of \source{} reported by \citet{2013Sci...340..950M}. The observations were carried out on 2011 August 25 and 2012 May 16 using the same phase reference and check sources as the VLBA observations. The observations were taken at a central frequency of 5.0\,GHz in dual polarisation mode, with 128\,MHz of bandwidth per polarisation. Data reduction was carried out in AIPS and flux densities were calculated by fitting a point source in the image plane (Table~\ref{tab:evn}).

\begin{table}
\caption{5\,GHz EVN observations of \source{} reported by \citet{2013Sci...340..950M}. Specified Modified Julian Dates (MJDs) are for the mid-point of the observation, with the quoted uncertainties reflecting the observation duration. Systematic flux uncertainties of 5\% (at 1$\sigma$) have not been included.}
\centering
\label{tab:evn}
\begin{tabular}{ccc}
\hline
Date & MJD & Flux density\\
& & (mJy)\\
\hline

2011 Aug 25 & 55798.95$\pm$0.12 & 3.05$\pm$0.11 \\ 
2012 May 16 & 56063.17$\pm$0.12 & 1.29$\pm$0.05 \\ 

\hline

\end{tabular}
\end{table}

\subsection{{\it RXTE}}

We observed \source{} with {\it RXTE} during its 2010 outburst. Following \citet{2008ApJ...685..436A}, we used the 16\,s time-resolution Standard 2 mode PCA data to calculate X-ray colours and intensities. Hard and soft colours were defined as the count rate ratios (9.7--16.0 keV/6.0--9.7 keV) and (3.5--6.0 keV/2.0--3.5 keV), respectively, and intensity was defined as the 2.0--16.0\,keV count rate. Standard modelled background was subtracted (using PCABACKEST built in HEASOFT V6.9) and deadtime corrections were made on a per-PCU, per instrumental gain epoch basis. In the observations during which \source{} was the faintest, we detected the source at $\sim1.5$\,cts\,s$^{-1}$\,PCU$^{-1}$. We estimate any systematic error in the intensity (due to diffuse non-modelled background) to contribute $\sim0.5$\,cts\,s$^{-1}$\,PCU$^{-1}$ at its maximum.

\begin{figure*} 
\centering
\includegraphics[width=\textwidth]{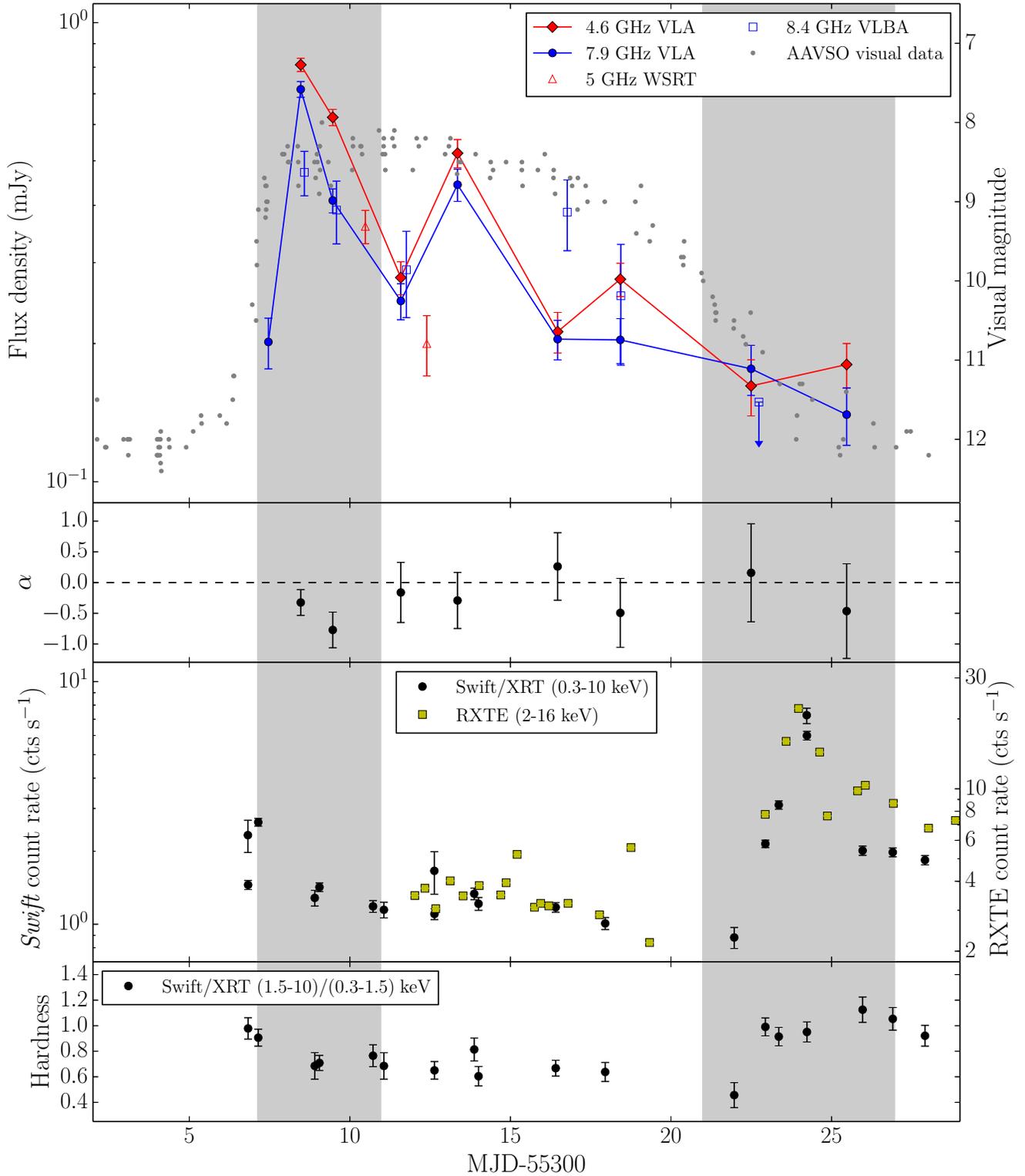}
\caption{Light curves of \source{} during the 2010 April outburst. Top panel: radio lightcurves from the VLA (filled diamonds and circles), VLBA (open squares) and WSRT (open triangles), where red and blue indicate $\sim$5\,GHz and $\sim$8\,GHz radio observations, respectively. Visual magnitudes from AAVSO are shown in light grey points, marking the progress of the outburst. Second panel: radio spectral index ($\alpha$, where $S_{\nu} \propto \nu^{\alpha}$). Third panel: \Swift/XRT and {\it RXTE} count rates, measured in the 0.3--10\,keV and 2--16\,keV bands, respectively. Bottom panel: \Swift/XRT hardness, defined as the count rate ratio (1.5--10)/(0.3--1.5) keV. Shaded panels represent the rise and decay periods of the outburst (see Section~\ref{sec:align} for definition), with the plateau phase in-between. The evolution of the 2010 outburst appears to follow the standard optical and X-ray outburst pattern for this source. }
\label{fig:lcs}
\end{figure*}

\subsection{{\it Swift}}

The \Swift{} X-ray telescope (\Swift/XRT) monitored \source{} during its 2010 April outburst. We retrieved the XRT observations from the High Energy Astrophysics Science Archive Research Centre (HEASARC) public archives. Observations of ~1\,ks were obtained on a roughly daily basis between April 20 and May 11, except for a 4-day interval between May 1 and May 5. All of the XRT data were collected in Photon Counting mode, and were extracted with the online XRT product generator \citep{2009MNRAS.397.1177E}, which corrects for the known pile-up. 

For the majority of the 2010 April outburst (for all times before May 9), the UV source was too bright to be usefully observed and saturated the UVOT. The final three observations collected data using the uvm2 filter, which has a central wavelength of 2246\,\AA. The magnitude at this time was uvm2 $\sim$11.75\,mag.

\section{Results}
\label{sec:results}

\subsection{Light curves}
\label{sec:lc}

The radio, optical and X-ray monitoring show the rise and decay of the 2010 April outburst of \source{} (Figure~\ref{fig:lcs}). Following the initial AAVSO detection, the optical monitoring showed a gradual rise for the first $\sim$1.5\,days of the outburst, followed by a sudden rapid increase. The optical emission then gradually declined for $\sim$10\,days, before a more rapid decline at the end of the outburst. 

\begin{figure}
\centering
\includegraphics[width=\columnwidth]{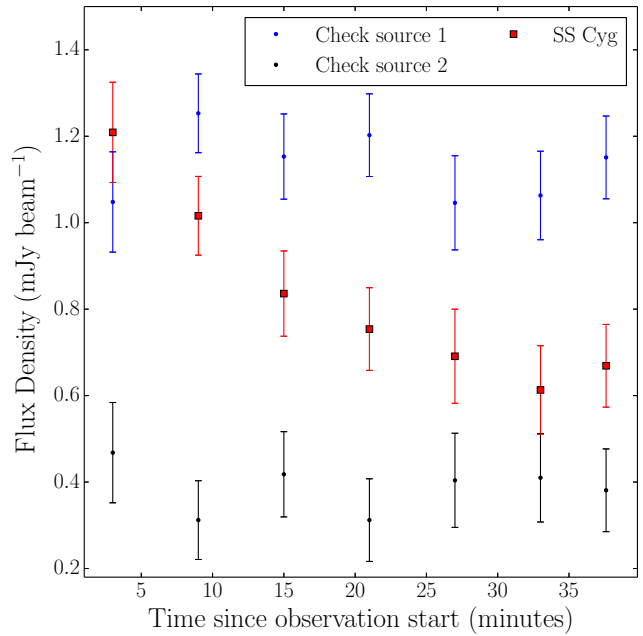}
\caption{Variability of \source{} (red squares) during the MJD 55308 VLA observation (which was the brightest radio observation of the 2010 outburst). Here we show the 5\,GHz observation binned into 6-minute time intervals. Check sources within the field (blue and black points) only varied within the noise, while \source{} varied significantly over the same time (according to statistical tests outlined by \citealt{2015MNRAS.450.4221B}).}
\label{fig:variabilityVLA}
\end{figure}

The radio emission brightened rapidly at the beginning of the outburst, with a detected peak of 0.81$\pm$0.03\,mJy at 4.6\,GHz $\sim$0.5\,days after the peak optical brightness (although variability within the observation showed a peak observed radio flux of 1.2$\pm$0.1\,mJy\perbeam\ at the start of the observation; Section~\ref{sec:fast_var} and Figure~\ref{fig:variabilityVLA}). During the rapid brightening, the radio emission evolved from a flat or slightly steep spectrum ($\alpha$=$-$0.3$\pm$0.2) to steep spectrum ($\alpha$=$-$0.8$\pm$0.3) at the peak of the observed radio emission (Figure~\ref{fig:lcs}, second panel; although the uncertainties are large and the results are consistent at 1-sigma), before becoming consistent with flat ($\alpha$$\sim$0) as the outburst began to fade. The radio emission brightened $\sim$3 days later, before fading. However, our VLBA monitoring showed an increase in flux on MJD 55316, that was discrepant from our VLA data that could be a flare. Although, this measurement is only $\sim$2-$\sigma$ from the VLA points. During the decay period, large uncertainties do not allow us to place strong constraints of the radio spectrum (which could be consistent with either a flat or steep spectrum). 

The X-ray monitoring showed the object brighten slightly at similar times to the optical before the X-ray flux dropped and remained steady during the optical peak of the outburst. During this steady plateau phase the X-ray spectrum was at it softest. The X-ray flux once again increased as the optical emission rapidly faded (Figure~\ref{fig:lcs}), and the X-ray spectrum hardened.

\begin{figure}
\centering
\includegraphics[width=\columnwidth]{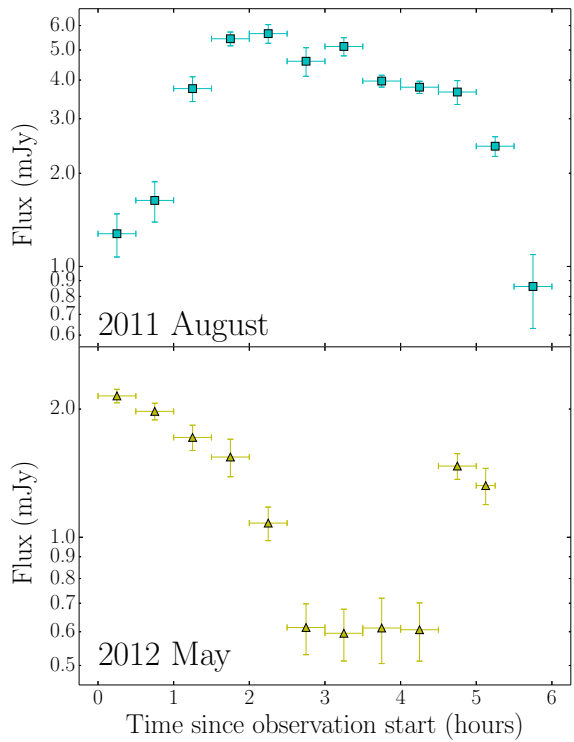}
\caption{Variability of \source{} during the 2011 August (top panel) and 2012 May (lower panel) EVN observations, which were both taken near the start of the outburst. Binning the 5\,GHz observation into 30-minute time intervals shows that the target varied by up to 5\,mJy and 1.5\,mJy over a few hours during the 2011 August and 2012 May EVN observations, respectively.}
\label{fig:variabilityEVN}
\end{figure}

\subsubsection{Rapid radio variability}
\label{sec:fast_var}

We searched for variability within each individual VLA observation, finding evidence for significant changes in the source flux density only on MJD 55308 (near the peak of the outburst), following statistical tests outlined in section\,4 of \citet{2015MNRAS.450.4221B}, which determine the $\chi^2$ probability that a source remained constant over the observations. Binning the data into 6-minute intervals (approximately the length of a scan), we find that on MJD 55308 the 4.6\,GHz flux dropped from 1.2$\pm$0.1 to 0.6$\pm$0.1\,mJy\perbeam\ during the observation (approximately 40\,minutes; Figure~\ref{fig:variabilityVLA}), indicating that the peak of the radio flux was at least 1.2\,mJy\perbeam. The corresponding 7.9\,GHz flux also dropped from 0.97$\pm$0.08 to 0.48$\pm$0.07\,mJy\perbeam. Checks indicated that the other sources in the field varied only within the noise level in this time (Figure~\ref{fig:variabilityVLA}), and therefore we believe this variability to be real. The decrease in measured flux density may explain the discrepancy between the VLA and VLBA flux density measurements on this date (Figure~\ref{fig:lcs}). The first hour of the VLBA run was simultaneous with the VLA observation.  If the trend of decreasing flux density continued for the following 6\,h, the averaged flux detected by the VLBA would be lower than the VLA measurement, in agreement with the observations.

We do observe some radio variability during our observations on MJD 55309 and MJD 55311. However, the change was not statistically significant (following \citealt{2015MNRAS.450.4221B}), and therefore cannot be definitively confirmed. We do not detect variability within any other individual observation.

Binning the 2011 August and 2012 May EVN observations from \citet{2013Sci...340..950M}, which were taken near the start of the outburst, into 30-minute intervals also shows significant variability within each observation (Figure~\ref{fig:variabilityEVN}). During the 2011 EVN observation, \source{} brightened from 1.3$\pm$0.2\,mJy to 5.6$\pm$0.4\,mJy in $\sim$2\,hours, before fading to 0.9$\pm$0.2\,mJy by the end of the observation. Within the 2012 EVN observation, the source faded from 2.15$\pm$0.08\,mJy to 0.59$\pm$0.08\,mJy over $\sim$3\,hours before increasing again to $\sim$1.5\,mJy by the end of the observation ($\sim$2\,hours later).

\subsection{Imaging}
\label{sec:resolved}

\begin{figure}
\centering
\includegraphics[width=\columnwidth]{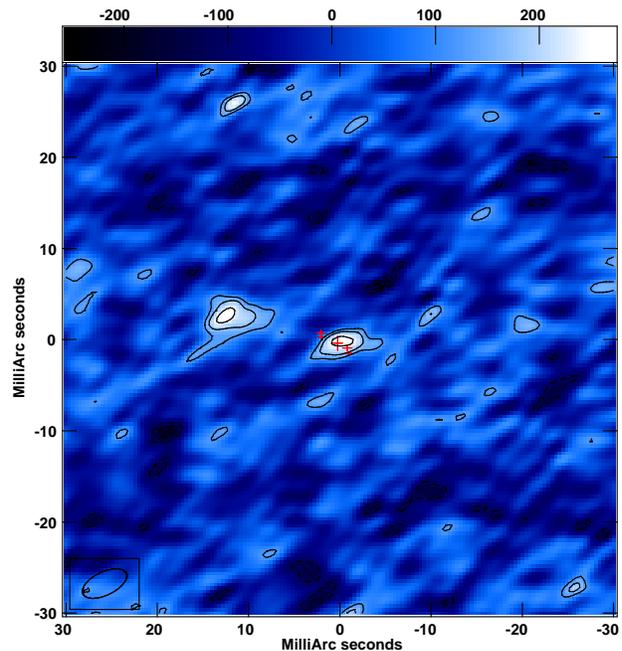}
\caption{Naturally-weighted VLBA image of \source{} on MJD 55311 (our third VLBA epoch). Contours are at levels of $\pm(\sqrt{2})^n$ times the rms noise of 65\,$\mu$Jy\perbeam, where $n=2,3,4,...$ Only the short baselines ($<$2328\,km) were used, and longer baselines were downweighted with a Gaussian tapering function of FWHM 152\,M$\lambda$. The colour bar at the top shows the image brightness in units of $\mu$Jy\perbeam, and the beam size is shown at bottom left. Red crosses show the core position over the five epochs. The crosses move from South-West to North-East with time owing to the proper motion (117.3$\pm$0.2\,mas\,yr$^{-1}$) and parallax signature of the system (see \citealt{2013Sci...340..950M}). The westernmost source is the core and the possible jet component is located to the east.}
\label{fig:jet}
\end{figure}

Resolving the radio emission with VLBI observations would provide strong evidence that the radio emission arises from a synchrotron-emitting jet.  We therefore searched for any hint of resolved radio emission in our VLBA monitoring epochs.  In four of the five epochs in which SS Cyg was detected with the VLBA, the source was unresolved, placing a constraint on the brightness temperature of the radio source, $T_{b}>$5.4$\times$10$^6$\,K, where
\begin{equation} 
T_{b}={\frac{c^2}{2 \nu^2 k_{\rm B}}\frac{S_{\nu}}{\Omega}},
\end{equation}
$c$ is the speed of light, $k_{\rm B}$ is Boltzmann's constant, and $\Omega$ is the solid beam angle. However, in the third epoch (MJD 55311), we found marginal evidence for a resolved jet at an angular separation of 12.1\,mas from the central source, along a position angle $70^{\circ}$\,E of N (Figure~\ref{fig:jet}). It was detected at a 4$\sigma$ level by only the shorter baselines to and between the southwestern antennas (KP, LA, OV, PT), and appears to have been resolved out on the longer VLBA baselines ($>$2328\,km). Its peak flux density was 240$\pm$60\,$\mu$Jy\,beam$^{-1}$, and the integrated flux density was 670$\pm$220\,$\mu$Jy.

To verify whether or not this component was real, we conducted a series of tests, splitting the data in two in the frequency, time, and polarization domains to determine whether the possible jet component was an artifact arising from instrumental or calibration issues in part of the data.  However, the low signal-to-noise ratio of the detection meant that such tests were inconclusive, with both the radio core and possible jet components dropping below the 3$\sigma$ level, as expected. We therefore tested removal of smaller fractions of the data, both in the time (removing 90-min chunks), frequency (removing single 8-MHz IF pairs), and baseline (removing single antennas) domains.  While the jet component was only detected when all four southwestern antennas were present, in no other case did it disappear completely. Furthermore, as expected from the low significance, gradually shifting the robust weighting scheme from natural to uniform made the jet component disappear as the noise level increased.  

The VLA flux density measured at 7.9\,GHz a few hours prior to the VLBA observations is lower than the summed flux of the core and possible jet components in the VLBA image (similar to that of the radio core alone, measured at 8.4\,GHz).  There is a tentative suggestion of variability seen within the 40-min VLA observation at this epoch, where the 7.9-GHz VLA flux density dropped by a factor of $\sim$2 during the observation (although tests showed this variability not to be significant; Section~{\ref{sec:fast_var}}), making it possible that the flux density could have changed by the time of the VLBA run.

In summary, we cannot convincingly determine whether or not this jet component is real from our data alone.  Future high-cadence observations with higher sensitivity on short to medium VLBI baselines (up to $\sim 2000$\,km) would be required to confirm or refute this suggestion of a jet.

If the component were to be real, we could use the baseline lengths on which it was resolved out to derive a size scale of 0.2--0.5\,AU for the ejecta.  At a projected angular separation of $\sim$1.4\,AU this would imply a jet opening angle in the range 8--20$^{\circ}$. The fact that the system was unresolved in both the preceding and subsequent VLBA observations (2 days before and 5 days after this observation) would suggest that either the ejecta are dark and we are seeing the hot spot where they impact the surrounding medium, or that the resolved component propagated out to an angular separation of 12\,mas in 2\,d, which would place a lower limit of $\sim$1200\,km\,s$^{-1}$ on the jet speed, well below the escape speed of a 0.81 M$_{\odot}$ white dwarf (the mass of \source{}; \citealt{2007ApJ...662..564B}). Inclination effects could increase this by a factor of $\lesssim$1.4 (\source{} is thought to have an inclination of 45--56$^\circ$; \citealt{2007ApJ...662..564B}). Transient jet emission from the highly-accreting neutron star system Scorpius X-1 \citep{1990ApJ...365..681H, 2001ApJ...558..283F} has been observed to brighten by a factor of $\sim$7 in less than an hour, and then fade by a factor of $\sim$2 over a few hours \citep{2001ApJ...558..283F}. Therefore, if dwarf novae launch jets in a similar way to their neutron star counterparts, shock brightening could account for the appearance of a resolved component in only one of our VLBA observations, while adiabatic expansion losses could then cause it to fade below detectability by the time of the subsequent observation. Thus, while the properties of the putative jet seem plausible, we will not consider it further in our discussion, pending a higher-significance detection from a future outburst.

\section{Discussion}
\label{sec:jd_coupling}

\subsection{The nature of the radio emission}

Following the arguments of \citet{2008Sci...320.1318K}, the observed radio emission may originate from free-free, synchrotron, or coherent emission processes. Free-free emission could be produced by a wind-formed gas cloud during an outburst \citep{1983PASP...95...69C, 1986A&A...154..377F}. At radio frequencies, optically-thick thermal free-free emission from an ionised gas generally produces brightness temperatures of 10$^4$--10$^5$\,K and a spectral index of $\sim$2 \citep{1979rpa..book.....R, 1992hea..book.....L}, inconsistent with our observed brightness temperature of $>$5.4$\times$10$^6$\,K and radio spectrum (Figure~\ref{fig:lcs}). The superposition of optically-thick and optically-thin free-free emission can produce a spectrum consistent with our observations. However, according to the disk instability model, the accretion rate of \source{} during outburst should be similar to its critical accretion rate of $\approx$1.4$\times$10$^{-8}$\,M$_{\odot}$\,yr$^{-1}$ \citep{2007A&A...473..897S}.

If we then assume that an un-collimated outflow carries all of the accreted material into an emitting cloud of ionised gas, the free-free radio flux can be expressed as \citep[][equation 8]{1975MNRAS.170...41W} 
\begin{equation}
S_{\nu}=23.2 \left( \frac{\dot{M}}{\mu v_{\infty}}  \right)^{4/3} \frac{\nu^{2/3}}{D^2} \gamma^{2/3} g^{2/3} Z^{4/3} \; {\rm Jy},
\label{eq:freefreeflux}
\end{equation}
where $\dot{M}$ is in $M_{\odot}$/yr, $D$ is the source distance in units of kpc, $v_{\infty}$ is the ejection velocity, $\mu$ is the mean atomic weight of the outflow, $g$ is the Gaunt factor, $\gamma$ is the ratio between the number of electrons and the number of ions, and $Z$ is the atomic charge. Assuming $\dot{M}=$1.4$\times$10$^{-8}$\,M$_\odot$\,yr$^{-1}$ \citep{2007A&A...473..897S}, $D$=0.114\,kpc \citep{2013Sci...340..950M}, $v_{\infty}$$\approx$6200\,km\,s$^{-1}$, which is the escape velocity of \source{}, assuming a white dwarf mass of 0.81\,M$_{\odot}$ \citep{2007ApJ...662..564B} and a radius of 5.5$\times$10$^8$\,cm, $\mu$$\approx$1.26, $g$$\approx$1.2, $\gamma$$\approx$1.26 and $Z$=1, we place an upper limit on the free-free radio flux of $\sim$2$\mu$Jy at 7.9\,GHz, much lower than our observed peak radio flare (of $\sim$1\,mJy at 7.9\,GHz during the 2010 monitoring). Even a clumpy wind with a filling factor of 1\% would not be able to produce the observed radio flux. Therefore, it is unlikely that the radio emission is of thermal origin. In fact, to produce the brightest observed 7.9\,GHz flux from the 2010 outburst, the mass accretion rate would need to be $\sim$2 orders of magnitude higher than the mass transfer rate predicted by the disk instability model.

Coherent emission may be produced by electrons trapped in the magnetosphere of the white dwarf \citep{1982ApJ...255L.107C}. This emission is characterised by a very steep radio spectrum with high levels of circular polarisation (close to 100\%). During the peak of the outburst (MJD 55308), we observe a relatively flat spectrum and place an upper limit on the circular polarisation of 3\% and 2.6\% at 4.6\,GHz and 7.9\,GHz, respectively. Gyrosynchrotron emission is able to produce the observed radio spectrum \citep{1989A&A...218..137B}. However, the low level of circular polarisation argues against this mechanism. Also, in agreement with \citealt{2008Sci...320.1318K}, for the brightness temperature to not exceed the Compton limit (of $\sim$10$^{12}$\,K) during the plateau phase of the outburst, the size of the emitting region must be larger than $\sim$75,000\,km. This is much larger than the expected magnetosphere size for a white dwarf (which is thought to be less than a few white dwarf radii; \citealt{2002MNRAS.335...84W}). \source{} is also thought to have a relatively low magnetic field \citep{2002MNRAS.335...84W}. Therefore, it is unlikely that the radio emission arises from magnetically-driven processes around the central white dwarf.

\begin{table*}
\caption{\source{} outburst parameters}
\centering
\label{tab:outbursts}
\begin{tabular}{cccccc}
\hline
Date & $t_{\rm rise}^{\rm a}$ & $t_{\rm decay}^{\rm a}$ & Duration$^{\rm b}$ & Start time$^{\rm b}$ & End time$^{\rm b}$\\
& (d\,mag$^{-1}$) & (d\,mag$^{-1}$) & (d) & (MJD) & (MJD) \\
\hline
1996 October & 0.61 & 2.39 & 6.09 & 50365.84 & 50371.93\\
2007 April   & 0.49 & 2.68 & 12.20 & 54215.04 & 54227.25\\
2010 April   & 0.56 & 2.50 & 13.86 & 55307.09  & 55320.95\\
2011 August  & 0.55 & 1.98 & 6.23 & 55797.80 & 55804.03 \\
2012 May     & 0.58 & 2.29 & 13.74 & 56062.13 & 56075.86 \\

\hline
\multicolumn{6}{l}{$^{\rm a}$Defined as half the crossing time between $m_{\rm vis}$=11 and $m_{\rm vis}$=9 \citep{1998ApJ...505..344C}.}\\
\multicolumn{6}{l}{$^{\rm b}$Defined as the crossing of $m_{\rm vis}$=10 \citep{1992ApJ...401..642C}}
\end{tabular}
\end{table*}

\begin{figure*}
\centering
\includegraphics[width=\textwidth]{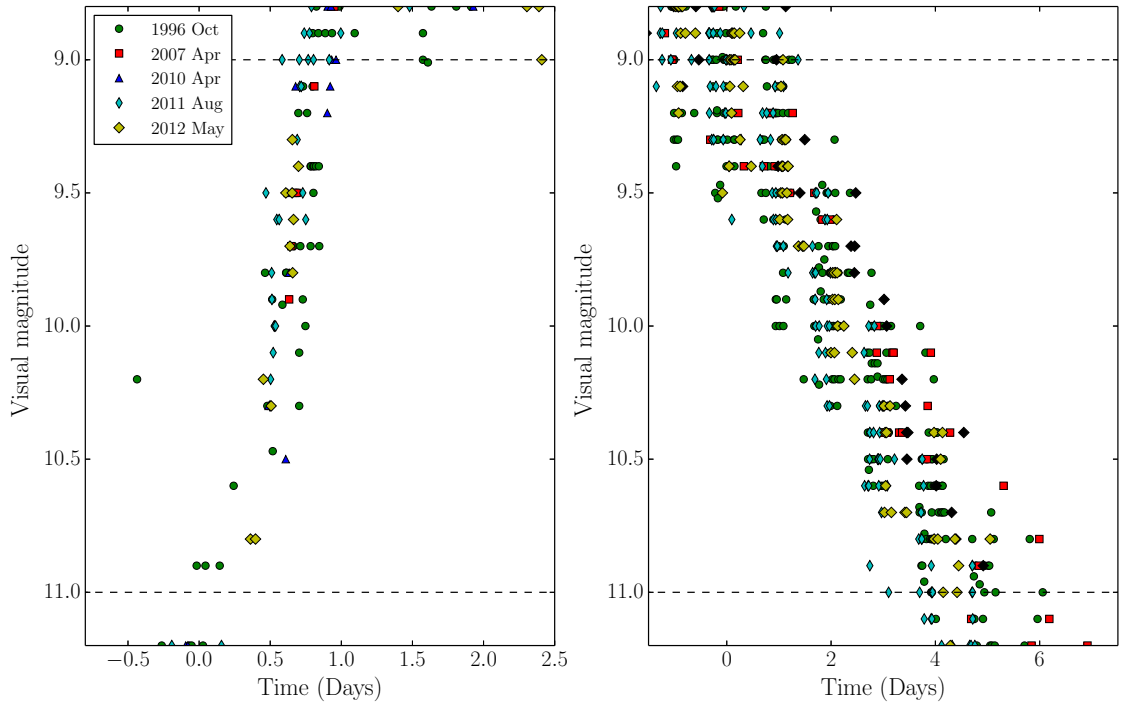}
\caption{The optical rise and decay of the 1996, 2007, 2010, 2011 and 2012 outbursts. Left panel: AAVSO visual band lightcurves of the rise, aligned to the interpolated m$_{\rm V}$=11 crossing. Right panel: AAVSO visual band lightcurves of the decay, aligned to the interpolated $m_{\rm vis}$=9 crossing times. The AAVSO magnitudes have a 0.1\,mag granularity, and the dashed horizontal lines show $m_{\rm vis}$=11 and $m_{\rm vis}$=9. The rise and decay times are consistent despite the bimodal outburst duration (Table~\ref{tab:outbursts}).}
\label{fig:risedecay}
\end{figure*}

Synchrotron emission is able to produce spectral indices between $-$1.5 and 2.5, as well as brightness temperatures up to 10$^{12}$\,K (if unbeamed) and is thus the most likely process responsible for the observed radio emission. If the tentative resolved jet component was real (Section~\ref{sec:resolved}), this would support the synchrotron nature of the radio emission. However, if the component was not real, we can still use our VLBI imaging to place constraints on the size scale of the jet. The VLBA observations recovered the full VLA flux density at all epochs except the first (where the discrepancy between the VLA and VLBA fluxes can be explained by the observed variability of the source; see Section \ref{sec:fast_var}), suggesting that we are not resolving out any radio emission. Therefore, the emitting region would be constrained to within the VLBA beam size of 2.6$\times$1.2\,mas$^2$. At a distance of 114\,pc \citep{2013Sci...340..950M}, this corresponds to a size scale of $<$0.3\,AU. For \source{} to have remained unresolved at the final VLBA epoch in which the source was detected ($\sim$11\,d after the onset of the outburst and assuming the resolved component was not real), the average expansion velocity would need to be $<$47\,km\,s$^{-1}$, more than two orders of magnitude below the escape velocity from the surface of the white dwarf, which seems implausible. Therefore, in this scenario, unless the radio emission can be attributed to the donor star, an unresolved compact jet seems the most likely explanation. Linear polarisation would help to diagnose the emission mechanism. However, polarisation calibration was not set up and even if it had been, at our sensitivity level we would only have been able to significantly detect linear polarisation (5$\sigma$) above 10\% in the first epoch.

\begin{figure*} 
\centering
\includegraphics[width=0.8\textwidth]{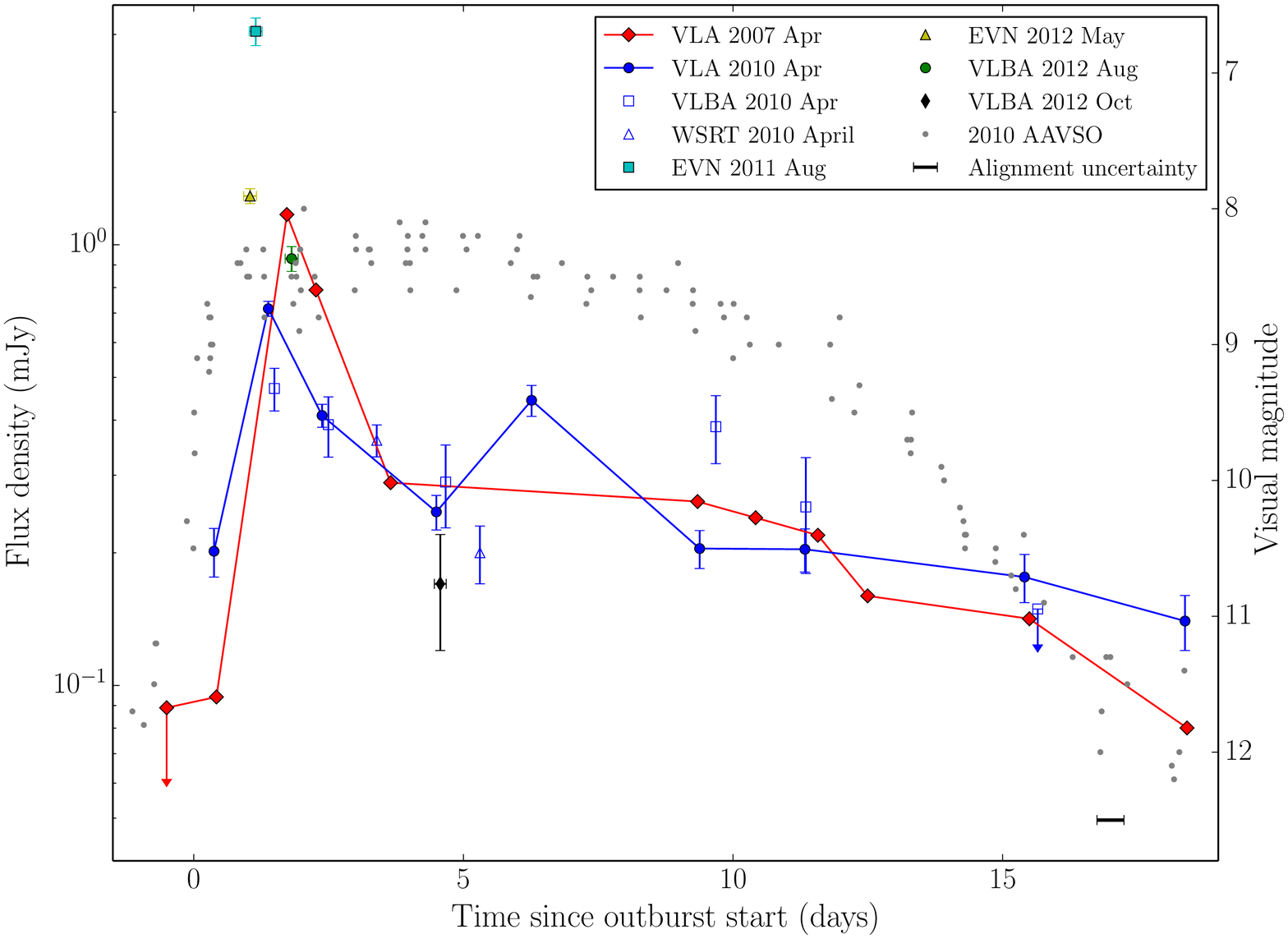}
\caption{The aligned radio observations of \source{} during outburst. The outburst start times are determined by the m$_{\rm vis}$=10 crossing times. Here we show the AAVSO (grey dots) and radio (blue points) monitoring of the 2010 April outburst, the 8.6\,GHz VLA radio observations (red diamonds) presented in \citet{2008Sci...320.1318K} and the 8.4\,GHz VLBA and 5.0\,GHz EVN observations from \citet{2013Sci...340..950M}, showing the averaged 2011 August (cyan square) and 2012 May (yellow triangle) EVN observations, as well as the averaged 2012 August (green circle) and 2012 October (black triangle) VLBA observations. The black error bar in the bottom right shows the $\pm$0.25\,day error on the alignment (where the uncertainty arises from the optical scatter). Here we see the similarity between the 2007 and 2010 radio lightcurves, as well as the similar timing of the initial bright radio flare across multiple outbursts.}
\label{fig:radio_outbursts}
\end{figure*}

\subsection{Outburst comparison}
\label{sec:align}
As suggested by \citet{2008Sci...320.1318K}, we can use the extreme ultraviolet emission of the source as a tracer for the luminosity arising from the accretion disk and boundary layer. Since the source was too bright for {\it Swift} UVOT, we were unable to obtain any coverage of the 2010 April outburst in the ultraviolet band, except at the very end of the outburst. The most complete ultraviolet coverage in the literature is the {\it EUVE} data published by \citet{2003MNRAS.345...49W}. Thanks to the detailed AAVSO time series over the past century, we are able to compare the dense multiwavelength monitoring of the 1996 October outburst with the 2007, 2010, 2011 and 2012 radio outbursts to estimate the UV and X-ray behaviour during the radio outbursts, under the assumption that the rise and decay behaviour is sufficiently similar between the events. This assumption is supported by the narrow distribution of rise and decay times of the outbursts of \source{} \citep{1998ApJ...505..344C}. Also, an analysis of the hard X-ray emission of four separate outbursts between 1996 and 2000 showed that the morphology was essentially the same, despite the different outburst durations \citep{2004ApJ...601.1100M}.

\begin{figure*} 
\centering
\includegraphics[width=\textwidth]{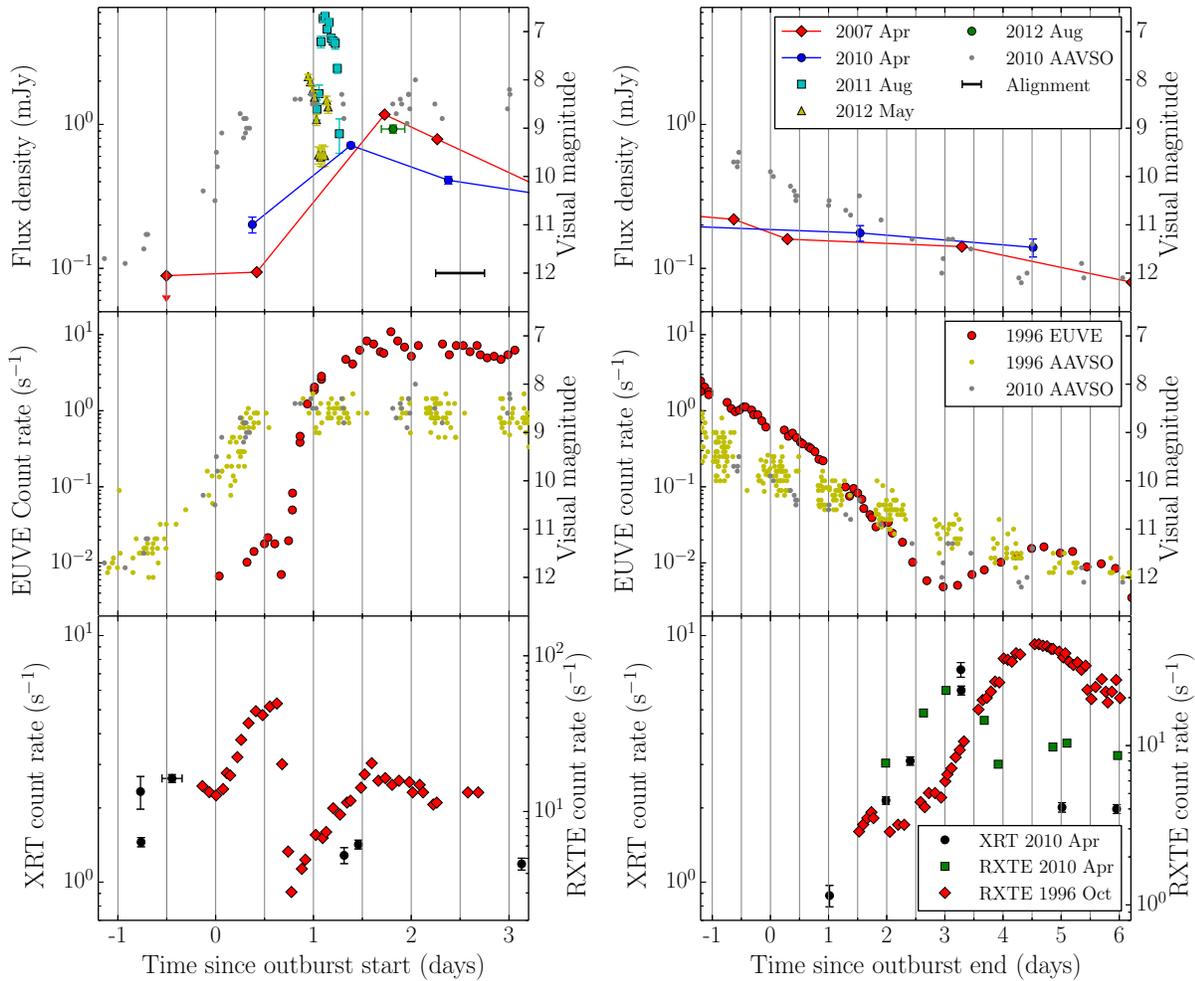}
\caption{Comparison of the rise (left panels) and decay phases (right panels) of multiple outbursts from \source{}. The outburst start and end times are aligned according to the interpolated m$_{\rm vis}$=10 crossing times (Section~\ref{sec:align}). The $\pm$0.25\,day error in the alignment is shown as the black error bar at the bottom of the top left panel and half-day intervals are shown by the vertical grey lines. The top panels show the AAVSO (grey dots) and 7.9\,GHz radio (blue points) monitoring of the 2010 April outburst, the 8.6\,GHz VLA radio observations of the 2007 outburst (red diamonds; \citealt{2008Sci...320.1318K}), as well as the 5.0\,GHz EVN and 8.4\,GHz VLBA observations, showing the 2011 August (cyan squares), 2012 May (yellow triangles) and 2012 August (green points) observations \citep[taken from][]{2013Sci...340..950M}. The second panels link the 2010 April AAVSO observations (grey points) to the EUVE (red points) and AAVSO (yellow points) monitoring of the 1996 October outburst from \citet{2003MNRAS.345...49W}. The bottom panels show the {\it Swift} and {\it RXTE} lightcurves of the 2010 April outburst (black and green points) and the 1996 October {\it RXTE} lightcurve \citep[red squares; from][]{2003MNRAS.345...49W}. The similarities between the rise and decay of the optical outbursts of \source{} are well documented \citep[e.g.][]{1992ApJ...401..642C}, allowing us to link the emission at different wavebands. The monitoring shows that the radio emission switches on during the initial X-ray brightening, before peaking at similar times to the UV band. Following the initial flare the radio emission then shows a plateau phase for $\sim$5\,days before fading.}
\label{fig:outburst_comparison}
\end{figure*}

To align each outburst, we define the start and end time of the outburst as the weighted mean of the $m_{\rm vis}$=10 crossing time \citep{1992ApJ...401..642C}, although the source can be identified to be in outburst well before this magnitude is reached. We then define the rise and decay phases as the linear interpolation of the re-binned (to 1\,d) $m_{\rm vis}$=11 and $m_{\rm vis}$=9 crossing times \citep{1998ApJ...505..344C, 2003MNRAS.345...49W}. We find that despite the differences in duration (the 1996 and 2011 outbursts were short duration, while the 2007, 2010 and 2012 outbursts were long), the rise and decay phases of the outbursts compared well (Table~\ref{tab:outbursts} and Figure~\ref{fig:risedecay}), with the rise and decay times falling within 1- and 1.5-$\sigma$ of the mean determined by \citet{1998ApJ...505..344C}, respectively, justifying our comparison of the 2007, 2010, 2011 and 2012 outbursts to the 1996 event to estimate the UV and X-ray emission. Aligning the radio observations of multiple outbursts shows that, as at higher frequencies, the radio morphology appears to be consistent and reproducible during typical long and short duration outbursts (Figure~\ref{fig:radio_outbursts}), implying that the radio emission is likely to be driven by the same mechanism during each outburst. 

All outbursts show an initial bright flare ($\gtrsim$1\,mJy) within 0.5--2\,days of the $m_{\rm vis}$=10 crossing time. Binning the 2011 August EVN monitoring in 30-minute intervals shows that the 2011 radio outburst peaked at 5.6$\pm$0.4\,mJy 1.10$\pm$0.25\,days after the beginning of the outburst (where the uncertainty results from the optical scatter of the optical rise; Figure~\ref{fig:risedecay}). We also observe the fading stage of a similar flare in the 2012 outburst. While we only observe the decay of the 2012 May flare, the re-aligned 2011 and 2012 radio flare decays appear within 0.25\,days of each other, consistent within the optical scatter in the rise (of $\sim$0.5\,days), and therefore, could be coincident. 

The VLA monitoring also observed a second radio brightening (at a significance of $\sim$3--4$\sigma$) during the 2010 outburst approximately 6.5\,days after the beginning of the outburst (Figure~\ref{fig:radio_outbursts}). This flare (or others) was not detected during other outbursts. However, this could be due to a lack of radio monitoring at similar times during the other outbursts. Aside from this second flare, the radio outbursts appear to follow a similar decay, fading gradually, with the compact jet remaining on throughout (Figure~\ref{fig:radio_outbursts}).

\subsubsection{X-ray and UV outburst}

The optical and X-ray rise and decay light curves from the 2010 and 1996 outbursts show similar morphology (Figure~\ref{fig:outburst_comparison}), with the only difference being the timing of the X-ray rise observed at the end of the outbursts. This rise occurred $\sim$1.5\,days earlier in the 2010 outburst than it did in the 1996 one (Figure~\ref{fig:outburst_comparison}, right column, bottom panel), suggesting that the inward propagation of the cooling wave was not identical between outbursts, such that the state of the outer disk (as represented by the optical emission) does not correspond perfectly to a given set of conditions in the inner accretion flow (which is also shown by the 1--2\,day scatter between outbursts during the decline; Figure~\ref{fig:risedecay}). This shift in time is smaller than the spacing of our radio observations. Therefore, it should not significantly affect the predicted X-ray behaviour at the times of the radio observations.

At the start of the outburst, the initial X-ray rise began at a similar time as the $m_{\rm vis}$=10 crossing (Figure~\ref{fig:outburst_comparison}, left column, third panel), peaking $\sim$0.5 days after the outburst start before rapidly quenching. This rise is thought to occur as material first reaches the boundary layer, while the suppression results from the boundary layer becoming optically thick to its own emission \citep{2003MNRAS.345...49W}. At the same time as the X-ray suppression, the extreme-UV emission brightened dramatically. As discussed by \citet{2003MNRAS.345...49W}, the coincidence of the X-ray suppression and the extreme-UV rise indicates that these two components likely arise from the same emitting region (the boundary layer).

\subsection{Disc-jet coupling}
\label{sec:XRBsimilarities}
During the early stages of an outburst, radio emission is first detected at similar times to the initial X-ray flare, before brightening rapidly to $>$1\,mJy (up to $\sim$5.6\,mJy) within 0.5--1.8\,days of the outburst start (Figure~\ref{fig:outburst_comparison}, left column, first and second panels). The 2011 EVN monitoring shows the target brightening by $\sim$5\,mJy over $\sim$2\,hours, giving a variability brightness temperature of $\sim$5.5$\times$10$^3$\,K. The radio emission then exhibits a plateau phase for $\sim$5 days (although the 2010 monitoring shows a smaller second flare and subsequent flares may occur that were not sampled by the radio observations; Figure~\ref{fig:lcs}) and then fades. The initial X-ray flare is thought to occur as the disk material first reaches the boundary layer \citet{2003MNRAS.345...49W}, which is both vertically (along the white dwarf surface) and radially extended (with a dynamical width of $\sim$0.01\,R$_{\rm WD}$ and a thermal width over which the boundary layer luminosity is radiated of $\sim$0.1\,R$_{\rm WD}$; e.g. \citealt{1993Natur.362..820N, 1995ApJ...442..337P, 1997ASPC..121..230P, 2001ApJ...547..355P}). The coincidence of the radio switching on at the same time implies that the mechanisms responsible for the two could be related. Therefore, the presence of a boundary layer could be implicated in the generation of the observed radio emission in these objects. The high-resolution observations show sustained core radio emission from \source{} during the outburst (Figure~\ref{fig:lcs}) while the VLA monitoring showed a radio spectrum that was consistent with flat, indicating that the compact radio jet remained active throughout the observations.

\begin{table}
\caption{The range of radio and X-ray luminosities of \source{} during the flaring, plateau and decay phases of its 2010 outburst.}
\centering
\label{tab:lrlx}
\begin{tabular}{ccc}
\hline
Phase & $L_{\rm R}$ & $L_{\rm X}$ \\
 & 5\,GHz & 1--10\,keV \\
    & (erg s$^{-1}$) & (erg s$^{-1}$)  \\ 
\hline
Flaring & (0.4--4.7)$\times$10$^{26}$  & (0.5--3.7)$\times$10$^{32}$ \\
Plateau & (1.4--4.0)$\times$10$^{25}$ & (4.0--8.8)$\times$10$^{31}$  \\
Decay   & (1.0--1.5)$\times$10$^{25}$ & (0.1--3.7)$\times$10$^{32}$  \\

\hline
\end{tabular}
\end{table}

\begin{figure}
\centering
\includegraphics[width=\columnwidth]{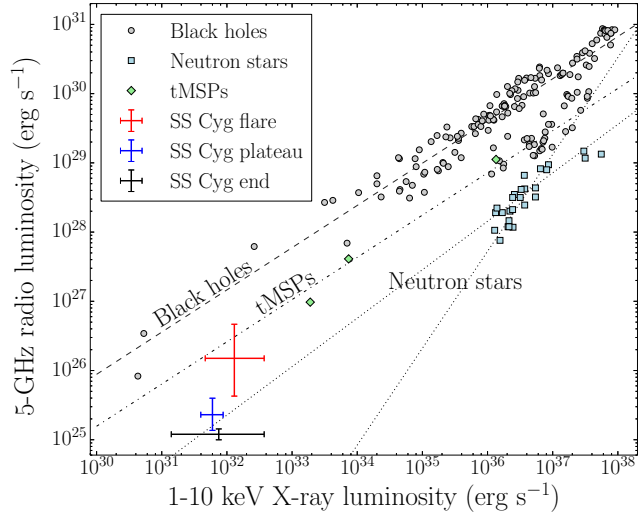}
\caption{The radio/X-ray correlation of \source{} during its initial flaring phase (red error bars), plateau phase (blue error bars) and decay phase (Figure~\ref{fig:lcs} and Section~\ref{sec:align}). We also plot representative black hole and neutron star systems, as well as transitional millisecond pulsars (tMSPs), with their proposed $L_{\rm R} \propto L_{\rm X}$ correlations. During its brightest radio flare (which only lasted a few hours; Figure~\ref{fig:variabilityEVN}), the $L_{\rm R}$/$L_{\rm X}$ of \source{} appeared as radio bright as the proposed tMSP track ($L_{\rm R} \propto L_{\rm X}^{\sim 0.6}$, shown by the dot-dashed line; \citealt{2015ApJ...809...13D}) but remained a factor of $\sim$4 below the expected radio luminosity of black holes at similar X-ray luminosities (shown by the dashed line). During the plateau and decay phase, \source{} was as radio bright as the $L_R \propto L_X^{\sim 0.7}$ \citep{2006MNRAS.366...79M} neutron star correlation extrapolated to low X-ray luminosities (shown by the shallower dotted line, where the steeper dotted line shows the $\sim$1.4 correlation; \citealt{2006MNRAS.366...79M}). Therefore, dwarf novae could be a source of confusion for quiescent neutron star systems when classifying them using radio/X-ray ratio, but not black holes.}
\label{fig:lrlx}
\end{figure}

Compact radio jets are observed from both black hole and neutron star X-ray binaries. While there are clear similarities between the two classes of object, there are important quantitative differences. For example, neutron star systems produce lower radio luminosities, $L_{\rm R}$ (defined as $4\pi D^2 \nu S_{\nu}$), at a given X-ray luminosity, $L_{\rm X}$, to their black hole counterparts \citep{2006MNRAS.366...79M} and the two classes display different correlations. Black holes generally exhibit a slope of 0.63$\pm$0.03 \citep[e.g.][]{2013MNRAS.428.2500C, 2012MNRAS.423..590G}, while neutron star systems are thought to display a slope of $\sim$0.7 for atoll and Z sources or $\sim$1.4 for hard state atoll sources \citep{2006MNRAS.366...79M}. Although transitional millisecond pulsars, tMSPs, may exhibit a similar $L_{\rm R}$ to $L_{\rm X}$ relationship as black hole systems; \citealt{2015ApJ...809...13D}. Recent works have used an object's behaviour in the $L_{\rm R}$/$L_{\rm X}$ plane to classify its nature \citep[e.g.][]{2013ApJ...777...69C, 2015MNRAS.453.3918M}. Therefore, although X-ray observations of dwarf novae do not sample the full accretion luminosity and cannot be used to compare the emission mechanisms between CVs and other objects, determining the $L_{\rm R}$/$L_{\rm X}$ ratio is important to understanding whether CVs could be a source of confusion when classifying black hole and neutron star X-ray binaries.

During its initial radio flaring phase, \source{} brightened to a radio luminosity of $\approx$4.4$\times$10$^{26}$\,erg\,s$^{-1}$ at a corresponding X-ray luminosity of $\approx$1$\times$10$^{32}$\,erg\,s$^{-1}$ (Table~\ref{tab:lrlx}), which is a factor of $\sim$4 below the black hole correlation at similar X-ray luminosities (Figure~\ref{fig:lrlx}). While this radio bright phase could produce a similar radio/X-ray correlation to the proposed tMSP relationship, the initial bright radio flare only lasted $\sim$4\,hours (Figure~\ref{fig:variabilityEVN}) before rapidly fading to around a factor of $\sim$4 less radio bright than the proposed tMSP radio luminosity at that X-ray luminosity (in $\sim$2 hours). During the plateau and decay phase of its outburst (which are more analogous to the low/hard states of their black hole and neutron star counterparts), \source{} showed a much higher radio luminosity than the widely adopted $L_R \propto L_X^{\sim1.4}$ neutron star X-ray binary correlation \citep{2006MNRAS.366...79M}. However, the observed luminosities could appear to be consistent with an $L_R \propto L_X^{\sim 0.7}$ neutron star correlation. These results show that while a flaring dwarf novae could not masquerade as a black hole, it could be a source of confusion to neutron star classification.

While black hole X-ray binaries exhibit strong quenching of the compact radio jet during their soft states (by up to 2.5 orders of magnitude; \citealt{1999ApJ...519L.165F, 2011MNRAS.414..677C, 2011ApJ...739L..19R}), at least two neutron star systems (4U 1820$-$30 and Ser X-1; \citealt{2004MNRAS.351..186M}) did not exhibit this same jet suppression in analogous accretion states (the neutron star system MXB 1730$-$335 may have also remained unquenched during outburst; \citealt{1998ATel....8....1R}). However, strong radio quenching has been observed in the neutron star systems Aquila X-1 \citep{2009MNRAS.400.2111T, 2010ApJ...716L.109M}, GX 9$+$9 \citep{2011IAUS..275..233M} and possibly 4U 1728$-$34 \citep{2003MNRAS.342L..67M}. Therefore, the lack of radio suppression in \source{} suggests a common (or similar) launching process between dwarf novae and those neutron star X-ray binaries that do not exhibit strong radio quenching. 

Our outburst comparison suggests that a vertically-extended region (i.e. the boundary layer) may be an important factor for the origin of the radio emission from \source{}. This result is consistent with the production of steady radio jets in both neutron star X-ray binaries and their black hole counterparts. Neutron star X-ray binaries possess a vertically-extended boundary layer, while black hole systems launch steady jets during their low/hard states \citep{2004MNRAS.355.1105F}, where the central regions of the inflow consist of a geometrically-thick radiatively inefficient accretion flow \citep{1995ApJ...452..710N}. The observed jet suppression in black hole systems is then thought to occur when the accretion flow becomes geometrically-thin during their soft states, whereas the magnetic field may play a role in the observed radio quenching of some neutron star systems \citep{2011IAUS..275..233M}.

\section{Conclusions}
\label{sec:conclusions}

We have compared the evolution of multiple radio outbursts of \source, demonstrating that the radio behaviour is reproducible during typical outbursts (both long and short duration). Our multi-frequency and high-resolution observations favour synchrotron emission over free-free and coherent emission mechanisms. Comparing the unquenched radio emission with optical, extreme-UV and X-ray monitoring shows that the radio emission appears to be switched on as disk material first reaches the white dwarf boundary layer. These results indicate that the boundary layer may play a role in jet production in these systems. We observed rapid radio variability (on $\lesssim$30\,minute timescales) at the peak of its outbursts. Multiwavelength monitoring of \source{} showed radio luminosities well below what is expected for black holes at similar X-ray luminosities, but consistent with a $L_R \propto L_X^{\sim 0.7}$ neutron star radio/X-ray correlation extrapolated to low luminosities. Our high-resolution monitoring provides marginal evidence for a resolved jet in one VLBI epoch. However, further observations are required to confirm this result. Our results show that white dwarf systems are capable of launching jets and that a vertically extended flow may play a role in the jet launching process in these systems.

\section*{Acknowledgments}

This research was supported under the Australian Research Council's Discovery Projects funding scheme (project number DP 120102393). We acknowledge the AAVSO network of variable star observers for providing the optical lightcurves of \source{}. JCAM-J is the recipient of an Australian Research Council Future Fellowship (FT140101082). GRS acknowledges funding from an NSERC Discovery Grant. DA acknowledges support from the Royal Society. KLP acknowledges support from the UK Space Agency. This work was supported by the Spanish Ministerio de Econom\'{a} y Competitividad (MINECO) under grant  AYA2013-47447-C3-1-P (SM). This work has made use of NASA's Astrophysics Data System. This research made use of APLpy, an open-source plotting package for Python hosted at http://aplpy.github.com. The International Centre for Radio Astronomy Research is a joint venture between Curtin University and the University of Western Australia, funded by the state government of Western Australia and the joint venture partners. The National Radio Astronomy Observatory is a facility of the National Science Foundation operated under cooperative agreement by Associated Universities, Inc. The Westerbork Synthesis Radio Telescope is operated by the ASTRON (Netherlands Institute for Radio Astronomy) with support from the Netherlands Foundation for Scientific Research (NWO). The European VLBI Network is a joint facility of independent European, African, Asian, and North American radio astronomy institutes. This research has made use of data and/or software provided by the High Energy Astrophysics Science Archive Research Center (HEASARC), which is a service of the Astrophysics Science Division at NASA/GSFC and the High Energy Astrophysics Division of the Smithsonian Astrophysical Observatory.


\label{lastpage}

\end{document}